\documentclass[twocolumn,english,prl, twocolumn,raggedbottom,showpacs, shortbibliography,floatfix, superscriptaddress]{revtex4}
\usepackage[T1]{fontenc}
\usepackage[utf8]{inputenc}
\usepackage{babel}
\usepackage{verbatim}
\usepackage{amsmath}
\usepackage{amssymb}
\usepackage{graphicx}
\usepackage{esint}
\usepackage[unicode=true,
 bookmarks=false,
 breaklinks=false,pdfborder={0 0 1},colorlinks=false]
 {hyperref}

\makeatletter
\@ifundefined{textcolor}{}
{%
 \definecolor{BLACK}{gray}{0}
 \definecolor{WHITE}{gray}{1}
 \definecolor{RED}{rgb}{1,0,0}
 \definecolor{GREEN}{rgb}{0,1,0}
 \definecolor{BLUE}{rgb}{0,0,1}
 \definecolor{CYAN}{cmyk}{1,0,0,0}
 \definecolor{MAGENTA}{cmyk}{0,1,0,0}
 \definecolor{YELLOW}{cmyk}{0,0,1,0}
}

\usepackage{graphicx}
\usepackage{color}
\usepackage{babel}
\usepackage{soul}

\@ifundefined{textcolor}{}{%
 \definecolor{BLACK}{gray}{0}
 \definecolor{WHITE}{gray}{1}
 \definecolor{RED}{rgb}{1,0,0}
 \definecolor{GREEN}{rgb}{0,1,0}
 \definecolor{BLUE}{rgb}{0,0,1}
 \definecolor{CYAN}{cmyk}{1,0,0,0}
 \definecolor{MAGENTA}{cmyk}{0,1,0,0}
 \definecolor{YELLOW}{cmyk}{0,0,1,0}
}

\@ifundefined{definecolor}{color}{}
\usepackage{babel}

\makeatother

\begin{document}

\title{Topological Kondo insulators in one dimension: Continuous Haldane-type
ground-state evolution from the strongly-interacting to the non-interacting
limit }

\author{Franco T. Lisandrini}

\affiliation{Facultad de Ciencias Exactas Ingenier\'ia y Agrimensura, Universidad
Nacional de Rosario, Argentina}

\affiliation{Instituto de F\'isica Rosario, CCT-Rosario (CONICET-UNR), Bv. 27
de Febrero 210 bis, 2000 Rosario, Argentina}

\author{Alejandro M. Lobos}

\email{lobos@ifir-conicet.gov.ar}

\selectlanguage{english}%

\affiliation{Instituto de F\'isica Rosario, CCT-Rosario (CONICET-UNR), Bv. 27
de Febrero 210 bis, 2000 Rosario, Argentina}

\affiliation{Facultad de Ciencias Exactas y Naturales, Universidad Nacional de
Cuyo, 5500 Mendoza, Argentina }

\author{Ariel O. Dobry}

\affiliation{Facultad de Ciencias Exactas Ingenier\'ia y Agrimensura, Universidad
Nacional de Rosario, Argentina}

\affiliation{Instituto de F\'isica Rosario, CCT-Rosario (CONICET-UNR), Bv. 27
de Febrero 210 bis, 2000 Rosario, Argentina}

\author{Claudio J. Gazza}

\affiliation{Facultad de Ciencias Exactas Ingenier\'ia y Agrimensura, Universidad
Nacional de Rosario, Argentina}

\affiliation{Instituto de F\'isica Rosario, CCT-Rosario (CONICET-UNR), Bv. 27
de Febrero 210 bis, 2000 Rosario, Argentina}
\begin{abstract}
We study, by means of the density-matrix renormalization group (DMRG)
technique, the evolution of the ground state in a one-dimensional
topological insulator, from the non-interacting to the strongly-interacting
limit, where the system can be mapped onto a topological Kondo-insulator
model. %
{} We focus on a toy model Hamiltonian (i.e., the interacting ``$sp$-ladder''
model), which could be experimentally realized in optical lattices
with higher orbitals loaded with ultra-cold fermionic atoms. Our goal
is to shed light on the emergence of the strongly-interacting ground
state and its topological classification as the Hubbard-$U$ interaction
parameter of the model is increased. Our numerical results show that
the ground state can be generically classified as a symmetry-protected
topological phase of the Haldane-type, even in the non-interacting
case $U=0$ where the system can be additionally classified as a time-reversal
$\mathbb{Z}_{2}$-topological insulator, and evolves adiabatically
between the non-interacting and strongly interacting limits. 
\end{abstract}

\date{\today}


\maketitle
\textbf{\textit{Introduction.}} The search for novel states of matter
with nontrivial topology has become an exciting pursuit in condensed
matter physics \citep{Kane05_Z2_Topological_invariant_in_QSHE,Bernevig_book_TI_TSC,Hasan10_Topological_insulators_review,Qi11_Review_TI_and_TSC},
and have opened a promising new path towards fault-tolerant quantum
computation \citep{Nayak08_RMP_Topological_quantum_computation}.
An important theoretical progress has been made in recent years with
a complete classification of insulating/superconducting topological
free-fermion systems in terms of dimensionality and global symmetries
of the Hamiltonian \citep{Altland97_Symmetry_classes,Kitaev_TI_classification,Ryu10_Topological_classification}.
However, a key open question is how to extend this classification
to interacting phases%
. %
For instance, it has been shown recently that in one spatial dimension
(1D) interactions can completely modify the free-fermion topological
classification \citep{Fidkowski10_Effects_of_interactions_on_TIs,Turner11_Topological_phases_of_1D_fermions}.

Topological Kondo insulators (TKIs) are a particular class of strongly-interacting
heavy-fermion materials where this question naturally appears \citep{Dzero10_Topological_Kondo_Insulators,Dzero12_Theory_of_topological_Kondo_insulators,Dzero15_Review_TKIs}.
In Refs. \citep{Dzero10_Topological_Kondo_Insulators,Dzero12_Theory_of_topological_Kondo_insulators,Dzero15_Review_TKIs},
it was suggested that samarium hexaboride (SmB$_{6}$), a narrow-gap
Kondo insulator known for more than 40 years, should be reinterpreted
as a TKI, generating a great deal of excitement. However, despite
the existence of supporting evidence for the TKI scenario \citep{Wolgast13_SS_in_SmB6,Zhang13_SS_in_SmB6,Xu14_SARPES_in_SmB6,Kim14_Topological_SS_in_SmB6},
the nature of the insulating state of SmB$_{6}$ has recently been
put under debate \citep{Tan15_Unconventional_Fermi_surface_in_insulating_state,Wakeham16_Low-temperature_conducting_state_SmB6_and_Ce3Bi4Pt3,Laurita16_Anomalous_3D_conduction_within_the_Kondo_gap_of_SmB6,Biswas17_Suppression_of_magnetic_excitations_in_SmB6},
and more theoretical and experimental work is needed to fully understand
it. %

From a different perspective, atomic, molecular and optical (AMO)
systems have become an important tool to study strongly-correlated
topological phases, thanks to the recent progress in experimental
techniques. In particular, cold atoms loaded into higher-orbital optical
lattices with nontrivial band-structure topology have opened a new
front in the pursuit of novel phases of matter, both from the theoretical
\citep{Isaacson05_Multiflavor_Bose_Hubbard_in_optical_lattices,Kuklov06_Unconventional_Strongly_Interacting_BECs_in_Optical_Lattices,Sun11_Topological_semimetal_in_a_fermionic_optical_lattice,Li13_Topological_interacting_ladders,Chen16_Chern_Kondo_Insulator_in_an_Optical_Lattice}
and the experimental \citep{Muller07_State_Preparation_and_Dynamics_of_Ultracold_Atoms_in_Higher_Lattice_Orbitals,Wirth11_Superfluidity_in_p_band_optical_lattice,Soltan-Panahi11_QPT_in_multiorbital_optical_lattices,Olschlager11_Unconventional_Superfluid_Order_in_Higher_Bands_in_Optical_Lattices}
point of view. The topology of $p$-orbital bands combined with interaction
effects leads to interesting physical systems \citep{Lewenstein11_Orbital_dance},
among which a predicted ``Chern Kondo insulating'' phase in 2D optical
lattices \citep{Chen16_Chern_Kondo_Insulator_in_an_Optical_Lattice},
and robust topological states in strongly interacting ladder-like
optical lattices in 1D \citep{Li13_Topological_interacting_ladders},
could be relevant to understand the nature of the insulating state
of TKIs.%

As an attempt to better understand the properties of bulk TKIs, Alexandrov
and Coleman proposed recently a 1D toy-model for a TKI (i.e., a ``$p$-wave''
Kondo-Heisenberg lattice) \citep{Alexandrov2014_End_states_in_1DTKI}.
Analyzing the symmetries of this model in the mean-field approximation,
these authors argued that it should be classified as a class-D insulator,
according to the free-fermion topological classes \citep{Altland97_Symmetry_classes,Kitaev_TI_classification,Ryu10_Topological_classification}.
Soon after, using Abelian bosonization, density matrix renormalization
group (DMRG), and quantum Monte Carlo techniques, the ground state
of 1DTKIs was identified as a Haldane-type insulating phase \citep{Lobos15_1DTKI,Mezio15_Haldane_phase_in_1DTKIs,Hagymasi16_Characterization_of_TKI,Lisandrini16_Topological_phase_transition_in_1D_TKI,Zhong16_Hubbard_Anderson_TKI},
which has no evident connection to the non-interacting topological
classes of Refs. \citep{Altland97_Symmetry_classes,Kitaev_TI_classification,Ryu10_Topological_classification}.
This result came as a surprise, as the Haldane phase is a paradigmatic
topological phase in one dimension, originally associated to the spin-1
antiferromagnetic chain, and is the simplest example of a symmetry-protected
topological (SPT) phase \citep{affleck_klt_short,affleck_klt_long,kennedy_z2z2_haldane,Pollmann10_Entaglement_spectrum_of_topological_phases_in_1D,Pollmann12_SPT_phases_in_1D}.
More importantly, the results in Ref. \citep{Lobos15_1DTKI,Mezio15_Haldane_phase_in_1DTKIs,Hagymasi16_Characterization_of_TKI,Lisandrini16_Topological_phase_transition_in_1D_TKI,Zhong16_Hubbard_Anderson_TKI}
put in evidence the risk of using mean-field approaches to interpret
the properties of strongly-interacting topological phases. 

\begin{figure}[t]
\centering{}\includegraphics[clip,scale=0.25]{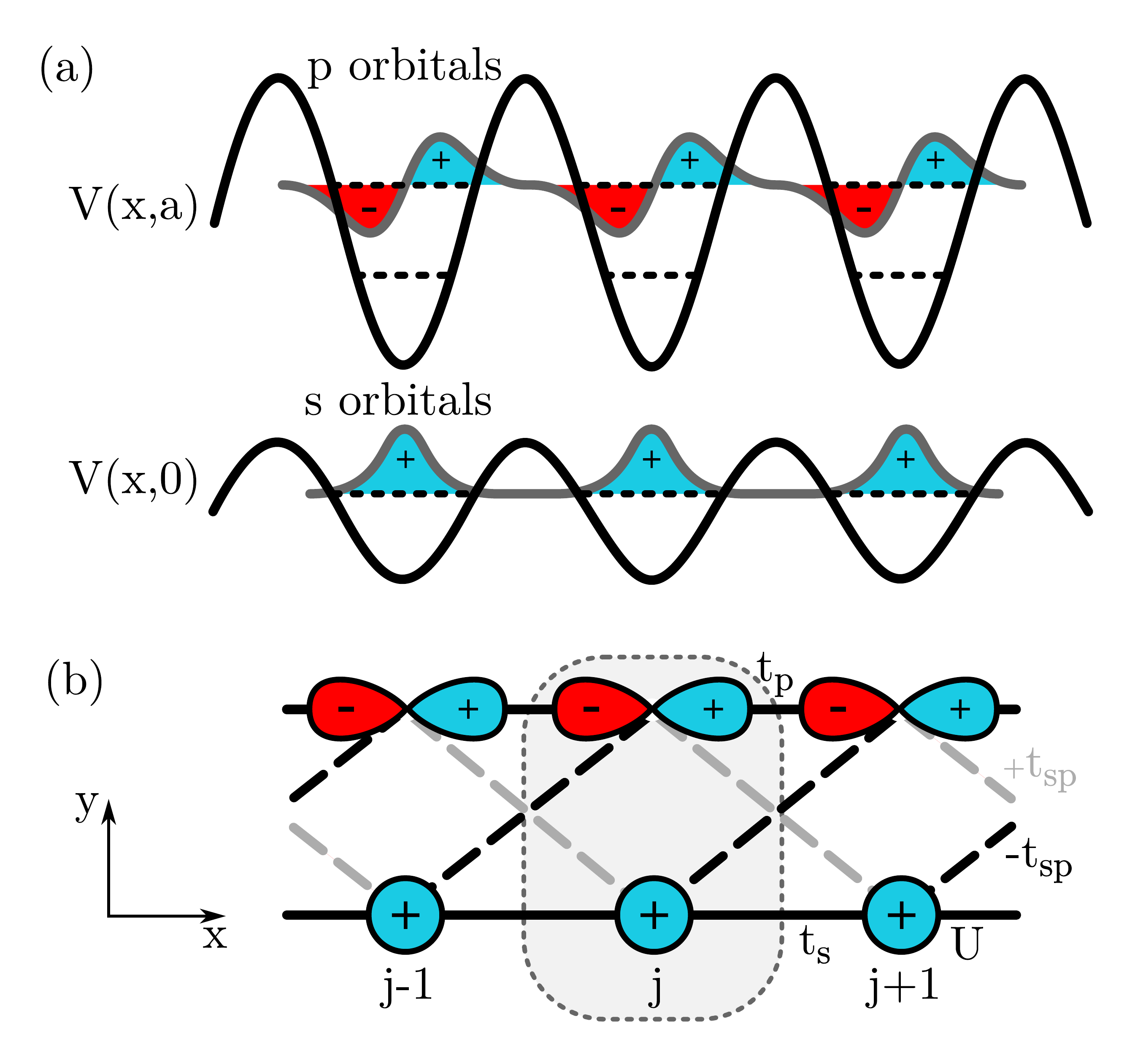}\protect\caption{\label{fig:spladder}(color online) (a) Scheme of a realization of
the $sp$-ladder model using the 2D laser potential $V\left(x,y\right)=V_{x}\sin^{2}\left(kx\right)+V_{1}\sin^{2}\left(ky\right)+V_{2}\sin^{2}\left(2ky+\phi\right)$
\citep{Li13_Topological_interacting_ladders}. Asymmetric double-well
potentials along the $\hat{y}$ axis, with $s$($p$) orbitals at
position $y=0$ ($y=a$), generate a $s$-($p$-) chain in the $\hat{x}$
direction. (b) Tight-binding approximation of $V\left(x,y\right)$,
where only the $s$ and $p$ orbitals are kept, generating the $sp$-ladder
structure.}
\end{figure}

Motivated by these developments, in this work we study, by means of
the DMRG technique, the effect of correlations on (an otherwise non-interacting)
1D $\mathbb{Z}_{2}$-topological band insulator, the 1D ``$sp-$ladder''
model (see Fig. \ref{fig:spladder}). Specifically, we study the evolution
of its ground state from the non-interacting to the strongly-interacting
limit, where the model can be mapped onto the ``$p$-wave'' Kondo-Heisenberg
model, and supports a Haldane-type ground state. We compute the entanglement
entropy, entanglement spectrum, and the string-order parameter of
the system, which are quantities typically used to characterize the
Haldane phase \citep{Pollmann10_Entaglement_spectrum_of_topological_phases_in_1D,Pollmann12_SPT_phases_in_1D},
as our interaction parameter (i.e., a local Hubbard-$U$ parameter)
is continuously varied. In addition, we compute the charge- and spin-excitation
gaps, and the charge and spin profiles of the topologically-protected
end states as a funcion of $U$. Our results show an \textit{adiabatic
evolution} of all ground-state properties as the interaction parameter
$U$ is increased, and the \textit{absence} of topological quantum
phase transitions (TQPT), suggesting that the ground state can be
generically classified as a symmetry-protected topological phase (SPT)
of the Haldane-type, \textit{even in the non-interacting case $U=0$}
where the system can be additionally classified as a time-reversal
$\mathbb{Z}_{2}$-topological insulator. This finding is surprising,
as these two ground states are usually considered to be qualitatively
different, and naively a TQPT separating these phases could be expected.
Our results are in stark contrast to the situation usually found in
higher-dimensional TKIs, where the interaction $U$ destabilizes the
non-interacting topological phase and induces gap-closing TQPTs at
critical values of $U$ separating topologically distinct phases \citep{Dzero10_Topological_Kondo_Insulators,Dzero12_Theory_of_topological_Kondo_insulators,Chen16_Chern_Kondo_Insulator_in_an_Optical_Lattice,Werner13_Interaction_driven_transition_in_TKIs,Legner14_Interaction_driven_TQPTs_in_TKIs}.
Moreover, our conclusions differ from recent related studies performed
on the same model using projective quantum Monte Carlo techniques
\citep{Zhong16_Hubbard_Anderson_TKI}, where the authors conclude
that the interaction $U$ breaks the adiabatic continuity of the groundstate,
despite the fact that their numerical results are consistent with
ours.

{} 

\textbf{\textit{Theoretical model. }}Let us consider, for concreteness,
ultracold fermions (for instance, $^{6}$Li or $^{40}$K) loaded into
the optical lattice of Fig. \ref{fig:spladder}(a). This situation
could be achieved by means of the 2D laser potential $V\left(x,y\right)=V_{x}\sin^{2}\left(kx\right)+V_{1}\sin^{2}\left(ky\right)+V_{2}\sin^{2}\left(2ky+\phi\right),$
for $V_{1},V_{2}\gg V_{x}$ \citep{Li13_Topological_interacting_ladders}.
In that limit, the system can be described as a collection of asymmetric
double-wells forming two-leg ladders in the $\hat{x}$ direction,
with relative depth of the two wells controlled by the phase $\phi$.
We will focus on the case where the $s$ orbitals in one leg and the
$p_{x}$ orbitals in the other leg are roughly degenerate. Assumming
orbitals well localized at the minima of the potential, we truncate
all other states in the description, and focus on the single-ladder
diagram of Fig. \ref{fig:spladder}(b). It can be seen that, due to
their different parity, the direct overlap between $s$ and $p_{x}$
orbitals on the same rung $j$ vanishes, i.e., $\left\langle s_{j\sigma}|p_{j\sigma}\right\rangle =0$,
while the overlap between next nearest-neighbor in different chains
has $p-$wave symmetry, i.e., $\left\langle s_{j\sigma}|p_{j+1\sigma}\right\rangle =-\left\langle s_{j\sigma}|p_{j-1\sigma}\right\rangle $
{[}hence the different sign of the inter-chain hopping $t_{sp}$ in
Fig.\ref{fig:spladder}(b){]}. 

We therefore model an interacting open-end $sp-$ladder with $L$
sites (i.e., rungs) as $H=H_{s}+H_{p}+H_{sp}+H_{U},$ where $H_{s}=t_{s}\sum_{j=1}^{L-1}\sum_{\sigma}\left(s_{j\sigma}^{\dagger}s_{j+1\sigma}+\text{H.c.}\right)$
and $H_{p}=-t_{p}\sum_{j=1}^{L-1}\sum_{\sigma}\left(p_{j\sigma}^{\dagger}p_{j+1\sigma}+\text{H.c.}\right)$
describe the legs of the ladder, which are connected via $H_{sp}=t_{sp}\sum_{j=1}^{L}\sum_{\sigma}\left[s_{j\sigma}^{\dagger}\left(p_{j+1\sigma}-p_{j-1\sigma}\right)+\text{H.c.}\right]$.
The operator $s_{j\sigma}$ ($p_{j\sigma}$) annihilates a fermion
with spin projection $\sigma$ in the $s$ ($p_{x}$) orbital at rung
$j$, and $n_{j\sigma}^{s\left(p\right)}$ is the corresponding number
operator. To account for the topological protected edge states, we
consider open boundary conditions. Finally, the Hubbard interaction
$H_{U}=U\sum_{j=1}^{L}\left(n_{j\uparrow}^{s}-\frac{1}{2}\right)\left(n_{j\downarrow}^{s}-\frac{1}{2}\right)$,
which can be physically produced and controlled in a cold-fermion
setup via Feshbach resonances, is assumed to exist only in the $s-$orbitals,
as we eventually want to make contact with the physics of strongly-interacting
Kondo insulators. %
{} %
This model is closely related to the (interacting) Shockely-Tamm \citep{Shockley39_Shockley_model,Tamm32_Tamm_model}
or Creutz-Hubbard ladder \citep{Creutz99_Creutz_ladder_model,Junemann16_Creutz_Hubbard_model}
models, and is such that when $U=0$ it describes a time-reversal
invariant $\mathbb{Z}_{2}$-topological band-insulator \citep{Kane05_Z2_Topological_invariant_in_QSHE,Zhong16_Hubbard_Anderson_TKI},
and when $U\gg t_{s},t_{sp}$ it can be mapped via a canonical transformation
onto the 1D $p-$wave Kondo-Heisenberg model, with Kondo and Heisenberg
exchange couplings $J_{K}=8t_{sp}^{2}/U$ and $J_{K}=4t_{s}^{2}/U$,
respectively, and where the ground state is a SPT phase of the Haldane
type \citep{Mezio15_Haldane_phase_in_1DTKIs}. In what follows, $t_{s}$
will be the unit of energies.

We now focus on the ground-state properties of $H$ at, or near to,
half-filling, i.e., when there are $N=2L$ fermions in the system.
For $U=0$, the topological features can be easily understood in the
``flat-band'' case $t_{s}=t_{p}=t_{sp}=t$, where $H$ can be written
in terms of the new fermionic operators $\gamma_{\pm,j\sigma}=\frac{1}{\sqrt{2}}\left(s_{j\sigma}\pm p_{j\sigma}\right)$
as $H_{\text{flat}}=2t\sum_{j=1,\sigma}^{L-1}\left(\gamma_{-,j\sigma}^{\dagger}\gamma_{+,j+1\sigma}+\text{H.c.}\right)$
(see Supplemental Material \ref{sec:supp_mat}). The topologically-protected
edge-modes in this representation are particularly simple, corresponding
to the operators $\gamma_{\alpha,\sigma}$ (with subindex $\alpha=\left\{ +,1\right\} $
or $\alpha=\left\{ -,L\right\} $), which drop from the Hamiltonian
and therefore correspond to zero-energy modes completely localized
at the ends \citep{Creutz99_Creutz_ladder_model,Li13_Topological_interacting_ladders,Continentino14_sp_ladders}.
In that case, we expect four degenerate states \textit{per end} (e.g.,
the states$\left|0\right\rangle _{\alpha},\left|\uparrow\right\rangle _{\alpha},\left|\downarrow\right\rangle _{\alpha},$
and $\left|\uparrow\downarrow\right\rangle _{\alpha}$ generated by
the application of $\gamma_{\alpha\uparrow}^{\dagger}$ and $\gamma_{\alpha\downarrow}^{\dagger}$).
When $U$ is turned on, this 4-fold degeneracy is locally split due
to the on-site repulsion, and the states $\left|0\right\rangle _{\alpha},\left|\uparrow\downarrow\right\rangle _{\alpha}$
become excited states while $\left|\uparrow\right\rangle _{\alpha},\left|\downarrow\right\rangle _{\alpha}$
span the ground state. 

{} 

{} 

\textbf{\textit{DMRG results}}. Following standard DMRG procedures
\citep{dmrg0}, we have implemented the Hamiltonian $H$ with open
boundary conditions, keeping $m=600$ states on every DMRG iteration,
which leads to a truncation error less than $O\left(10^{-9}\right)$
and to numerical errors in the subsequent figures much smaller than
symbol size. 

As shown in seminal works \citep{Turner11_Topological_phases_of_1D_fermions,Pollmann10_Entaglement_spectrum_of_topological_phases_in_1D,Pollmann12_SPT_phases_in_1D,Kitaev06_Topological_entanglement_entropy,Levin06_Topological_entanglement_entropy},
the entanglement entropy and entanglement spectrum can be used to
characterize SPT phases, such as the Haldane phase. In particular,
it was shown that the Haldane phase is characterized by an even-degenerate
entanglement spectrum \citep{Pollmann10_Entaglement_spectrum_of_topological_phases_in_1D}.
The entanglement spectrum is obtained calculating the eigenvalues
$\Lambda_{i}$ of the reduced density-matrix of the system $\hat{\rho}_{L/2}$,
obtained tracing out one half of the system, and the corresponding
entanglement (or von Neumann) entropy is $s\left(L/2\right)=-\text{Tr}\left\{ \hat{\rho}_{L/2}\ln\hat{\rho}_{L/2}\right\} =-\sum_{i}\Lambda_{i}\ln\Lambda_{i}$
\citep{Schollwoeck05_Review_DMRG}. A crucial observation is that
the degeneracy of the entanglement spectrum in a SPT phase cannot
change without a bulk phase transition, where the nature of the ground
state must change abruptly or where the correlation length must diverge
due to the closure of a bulk gap \citep{Turner11_Topological_phases_of_1D_fermions}.
Moreover, the occurrence of a gap-closing TQPT separating topologically
distinct phases should be accompanied by a logarithmic divergence
of the entanglement entropy $s\left(L/2\right)\sim k\ln\left(L\right)+\text{const}$
at the critical point \citep{Schollwoeck05_Review_DMRG}.

\begin{figure}
\centering{}\includegraphics[bb=20bp 10bp 420bp 288bp,scale=0.63]{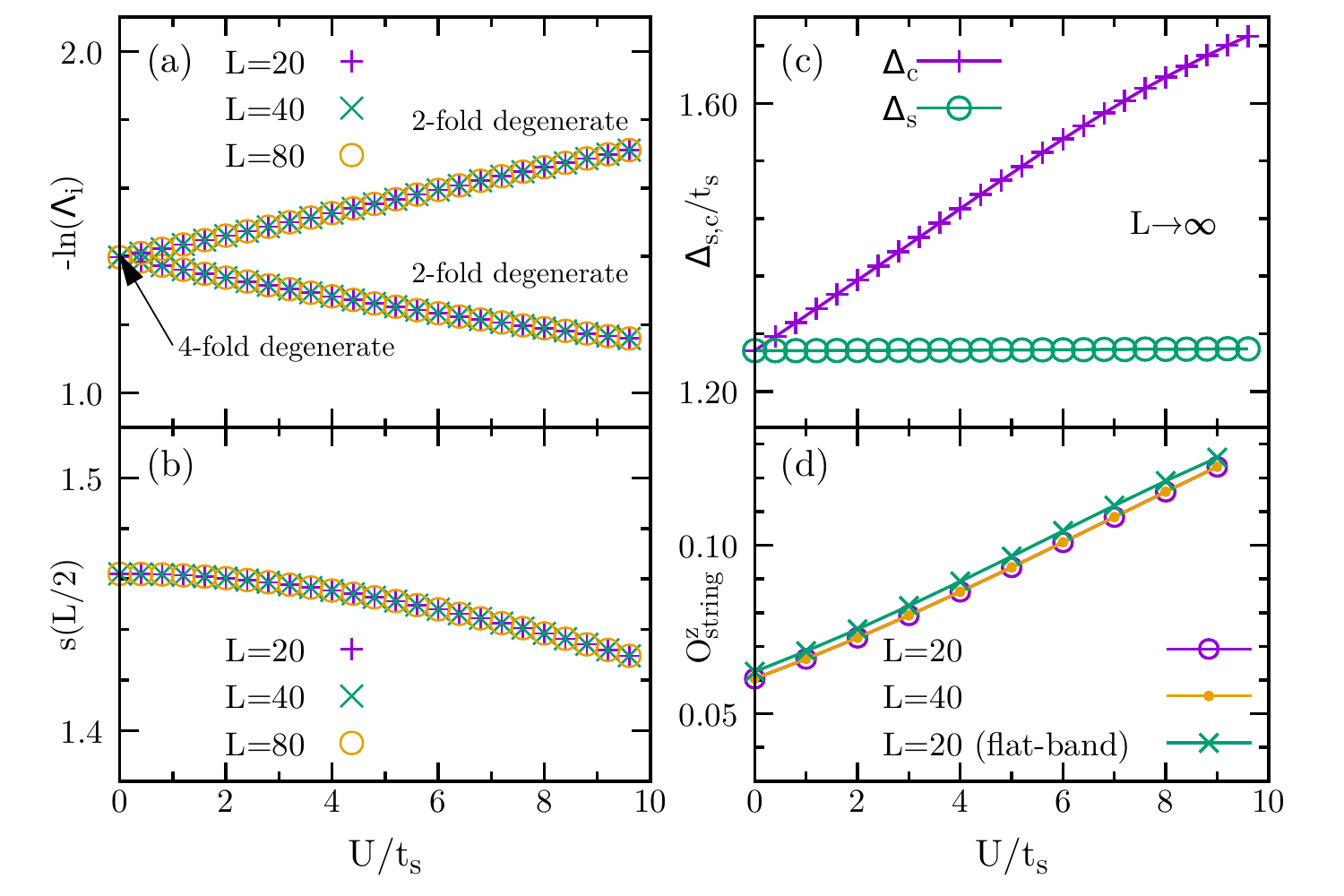}\protect\caption{(color online) (a) Largest eigenvalues $\Lambda_{i}$ of the entanglement
spectrum as a function of $U$, for different lattice sizes. (b) Entanglement
entropy of the reduced density matrix, $s\left(L/2\right)$, as function
of $U$ for different lattice sizes. (c) Spin and charge gap extrapolated
to the thermodynamic limit $L\rightarrow\infty$. (d) String order
parameter $\mathcal{O}_{\text{string}}^{z}$ as function of $U$ for
different system-sizes. \label{fig2}}
\end{figure}

In order to investigate for the existence of a bulk TQPT, we have
computed: (a) the entanglement spectrum, (b) the entanglement entropy,
(c) the charge- and spin-excitation gaps, and (d) the string order
parameter of $sp-$ladders ranging from $L=20$ to 80, total number
of fermions $N=\left(N^{\uparrow}+N^{\downarrow}\right)$, and total
spin projection $S^{z}=\frac{1}{2}\left(N^{\uparrow}-N^{\downarrow}\right)$.
Unless otherwise stated, we have used the parameters $t_{p}/t_{s}=\pi/10$,
$t_{sp}/t_{s}=1$, in order to recover the results obtained for the
$p$-wave Kondo-Heisenberg model studied in Ref. \citep{Mezio15_Haldane_phase_in_1DTKIs}
when $U\rightarrow\infty$. In Fig. \ref{fig2}(a) we show the evolution
of the largest eigenvalues $\Lambda_{i}$ of the entanglement spectrum
and in Fig. \ref{fig2}(b) the corresponding entropy $s\left(L/2\right)$.
The charge- and spin-excitation gaps in the 1D bulk, defined as $\ensuremath{\Delta_{\text{c}}\left(L\right)\equiv E_{0}\left(S^{z}=0,N=2L+4\right)-E_{0}\left(S^{z}=0,N=2L\right)}$
and $\ensuremath{\Delta_{\text{s}}\left(L\right)\equiv E_{0}\left(S^{z}=2,N=2L\right)-E_{0}\left(S^{z}=0,N=2L\right)}$,
respectively, where $E_{0}\left(S^{z},N\right)$ is the energy of
the ground state $\left|\psi_{0}\left(S^{z},N\right)\right\rangle $,
are shown in Fig. \ref{fig2}(c), where we have extracted the values
of $\Delta_{\text{c}}\left(L\right)$ and $\Delta_{\text{s}}\left(L\right)$
in the thermodynamic limit $L\rightarrow\infty$ by the means of a
finite-size scaling. 

All of our results show a smooth and continuous evolution as $U$
is varied, without any signature of a TQPT as the system size is increased.
In particular, the even-degenerate character of the entanglement spectrum
is preserved {[}see Fig. \ref{fig2}(a){]}, despite the continuous
breaking of the degeneracy existing at $U=0$ into two branches, presumably
corresponding to lower-lying spin and higher-lying charge degrees
of freedom. This is substantiated by the behavior of $\Delta_{\text{c}}\left(L\right)$
and $\Delta_{\text{s}}\left(L\right)$ shown in Fig. \ref{fig2}(c),
where the degeneracy of the charge and spin sectors at $U=0$ is broken
in a continuous fashion, and excitations become fractionalized into
low-lying spin and high-lying charge excitations. The entanglement
entropy $s\left(L/2\right)$, shown in Fig. \ref{fig2}(b), does not
present any divergence as the system size increases, and is bounded
from below by $s_{\text{min}}=\ln\left(2\right)\simeq0.693$, the
entropy of the Affleck-Kennedy-Lieb-Tasaki (AKLT) state corresponding
to the 2 degenerate end-states $\left|\uparrow\right\rangle ,\left|\downarrow\right\rangle $
\citep{Pollmann10_Entaglement_spectrum_of_topological_phases_in_1D}.
At $U=0$, a similar lower-bound is given by the flat-band state,
i.e., $s_{\text{min}}=\ln\left(4\right)\simeq1.386$, related to the
4 aforementioned degenerate end states.

In order to further clarify the crossover, we have calculated the
string correlator $\ensuremath{\mathcal{O}_{\text{string}}^{z}\left(l-m\right)=-\left\langle T_{l}^{z}e^{i\pi\sum_{l<j<m}T_{j}^{z}}T_{m}^{z}\right\rangle }$,
a non-local quantity that characterizes the breaking of the $Z_{2}\times Z_{2}$
hidden symmetry of the Haldane phase \citep{kennedy_z2z2_haldane}.
In Fig. \ref{fig2}(d) we plot the string order parameter $\mathcal{O}_{\text{string}}^{z}$
(i.e., the asymptotic value of the string correlator $\mathcal{O}_{\text{string}}^{z}\left(l-m\right)$
in the 1D bulk, taking $m=L/2$ and $l=L/2+d$ in order avoid the
effect of the boundaries) as a function of $U$. Again we see a smooth
and continuous adiabatic crossover between the non-interacting and
strongly-interacting limits. Strikingly, although this quantity has
been traditionally used to characterize strongly-interacting phases,
here we show that it yields\textit{ a finite value for all values
of $U$, even for $U=0$}, where the system has no net magnetic moments
due to strong charge fluctuations. This finite value of $\mathcal{O}_{\text{string}}^{z}$
for $U=0$ can be independently obtained by an analyical calculation
in the flat-band case, which yields the exact result $\ensuremath{\mathcal{O}_{\text{string}}^{z}\left(l-m\right)=1/16=0.0625}$
for $\left|l-m\right|\geqslant2$ (see Supplementary Material \ref{sec:supp_mat}).
The same value is obtained with DMRG with a relative error $O\left(10^{-13}\right)$.

\begin{figure}
\centering{}\includegraphics[bb=0bp 0bp 216bp 220bp,clip,scale=0.9]{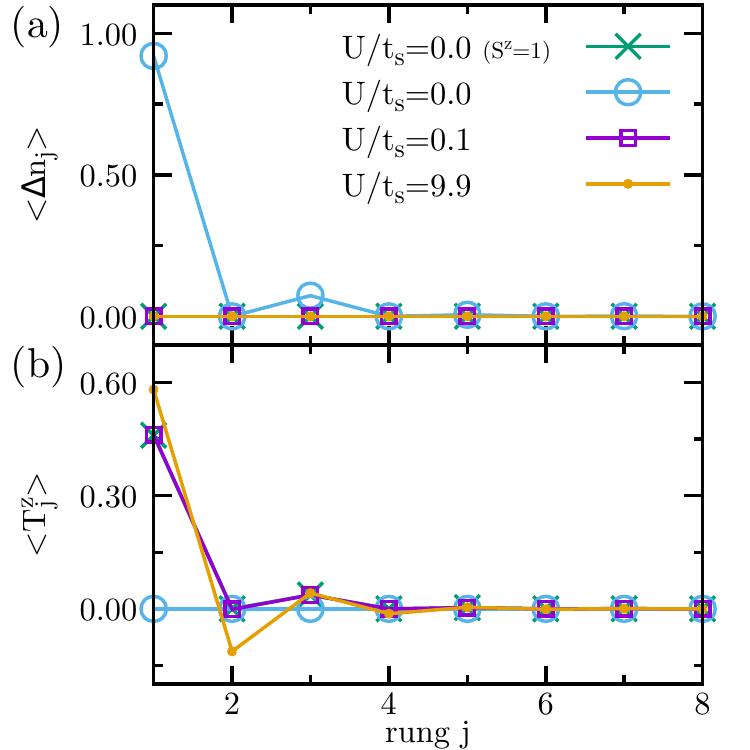}\protect\caption{(color online) (a) Charge and (b) spin profiles corresponding to a
ladder with $L=80$ sites and open boundary conditions, and for different
values of $U$. Only the left end is shown in both cases. All profiles
have been computed within the $S^{z}=0$ sector, except for $U=0$
where in addition we show the $S^{z}=+1$ profiles (represented by
green crosses).\label{fig3}}
\end{figure}

We now turn to the evolution of the topologically protected end-states.
We focus on the spatial spin profile, defined as $\left\langle T_{j}^{z}\right\rangle \equiv\langle\psi_{0}\left(S^{z},N\right)|T_{j}^{z}|\psi_{0}\left(S^{z},N\right)\rangle$,
where $\ensuremath{\mathbf{T}_{j}\equiv\mathbf{s}_{j}^{s}+}\mathbf{s}_{j}^{p}$,
is the total spin operator on the $j$-th rung, with $\mathbf{s}_{j}^{a}\equiv a_{j\mu}^{\dagger}\left(\frac{\boldsymbol{\sigma}_{\mu,\nu}}{2}\right)a_{j\nu}$
the spin-density of $a$-fermions ($a=\left\{ s,p\right\} $), and
$\boldsymbol{\sigma}_{\mu,\nu}$ the vector of Pauli matrices (summation
over repeated spin indices is implicit here). Invoking the $SU\left(2\right)$
spin-symmetry of the model, in what follows we will only compute the
numerically simpler $z$-component. Similarly, we define the (difference
in the) spatial charge profile as $\left\langle \Delta n_{j}\right\rangle \equiv\langle\psi_{0}\left(S^{z},N\right)|n_{j}|\psi_{0}\left(S^{z},N\right)\rangle-n_{0}$,
where $\ensuremath{n_{j}\equiv n_{j}^{s}+n_{j}^{p}}$ is the total
charge operator on rung $j$, and where we have substracted the uniform
background at half-filling $n_{0}=N/L=2$. In Fig.\ref{fig3}(a) and
(b) we plot $\left\langle T_{j}^{z}\right\rangle $ and $\left\langle \Delta n_{j}\right\rangle $
for different values of $U$ at half filling $N=2L$, and for different
$S^{z}$ subspaces compatible with this filling (i.e., $S^{z}=0$
and $S^{z}=\pm1$). Note that we only show the profiles at the left
end. Since the quantum numbers $S^{z}$ and $N$ must be specified
in the DMRG calculation, one must be careful in the interpretation
of the profiles at $U=0$, due to the 4-fold degeneracy. For $S^{z}=0$,
the DMRG procedure selects a ground state with a holon and a doublon
end-state (i.e., either $|0\rangle_{L}\otimes|\uparrow\downarrow\rangle_{R}$
or $|\uparrow\downarrow\rangle_{L}\otimes|0\rangle_{R}$), which has
a vanishing spin profile {[}see light blue circles in Figs. \ref{fig3}(a)
and (b){]}. However, for $S^{z}=+1$, the end-state combination $|\uparrow\rangle_{L}\otimes|\uparrow\rangle_{R}$,
which has free spin-1/2 end-states but a uniform charge profile {[}see
green crosses in Figs. \ref{fig3}(a) and (b){]}, is selected in the
ground state \citep{comment_sp_ladders}. However, our results are
fully consistent with the presence of free-fermion end-states of a
finite $\mathbb{Z}_{2}$ topological insulator. Interestingly, as
soon as the interaction $U$ is turned on (see violet square symbols
corresponding to $U/t_{s}=0.1$), due to the higher energy of charge
excitations, the charge profile ``freezes'' into the uniform configuration
$\left\langle n_{j}\right\rangle =n_{0}=2$, or equivalently $\left\langle \Delta n_{j}\right\rangle =0$,
which is physically consistent with the adiabatic emergence of a Mott-insulating
state at half-filling \citep{Lobos15_1DTKI,Mezio15_Haldane_phase_in_1DTKIs}.
The only surviving end-state at finite $U$ is therefore the spin
end-state, which continuously crosses-over into the usual Haldane
spin-1/2 end-states \citep{kennedy_z2z2_haldane}.

The adiabatic evolution of the ground state described here could be
ideally tested in ultracold $^{6}$Li or $^{40}$K gases, using recent
fermionic ``quantum-gas microscopes'', which allow to obtain the
full spin and charge density distributions with single-atom resolution
\citep{Haller15_Single-atom_imaging_of_fermions_in_a_quantum-gas_microscope,Cheuk15_Quantum-Gas_Microscope_for_Fermionic_Atoms,Cocchi16_Equation_of_State_of_the_2D_Hubbard_Model,Cheuk16_Observation_of_spatial_charge_and_spin_correlations_in_the_2D_Fermi-Hubbard_model,Parsons16_Site-resolved_measurement_of_the_spin-correlation_function_in_the_Fermi-Hubbard_model}.
Thanks to their high-resolution, such techniques have been used recently
to extract non-local correlation functions, such as the string-correlation
function, in balanced spin-mixtures of $^{6}$Li atoms trapped in
an asymmetric 2D optical lattice simulating a 1D Fermi-Hubbard model
\citep{Hilker17_Revealing_Hidden_AFM_Correlations_in_Doped_Hubbard_Chains_via_String_Correlators},
and could potentially be used to probe their evolution as the interaction
parameter is varied using Feshbach resonances. %

\textbf{\textit{Conclusions.}} Our work sheds light on the conceptually
relevant question of the classification of strongly-interacting topological
phases and their connection to the free-fermion topological classes.
Based on the even-degeneracy of the entanglement spectrum for all
values of the interaction $U$, we conclude that the ground state
of the system can be generically classified as a SPT Haldane-type
phase, \textit{even in the non-interacting case} $U=0$, where the
system additionally admits the usual classification as a non-interacting
time-reversal $\mathbb{Z}_{2}$-topological insulator \citep{Kane05_Z2_Topological_invariant_in_QSHE}.
We hope our work will trigger more theoretical and experimental research
in this direction. %

\textbf{\textit{Acknowledgements.}} F.T.L, A.O.D. and C.J.G. acknowledge
support from CONICET-PIP 11220120100389CO. A.M.L acknowledges support
from PICT-2015-0217 of ANPCyT. 

\bibliographystyle{apsrev_nourl}

\begin{thebibliography}{60}
\bibitem[{com()}]{comment_sp_ladders}
\bibinfo{note}{This is different to the results obtained in Figs. 4 and 5 in
  Ref. \cite{Zhong16_Hubbard_Anderson_TKI}, where a flat charge profile is
  obtained for $U=0$, presumably due to the statistical average performed by
  the quantum Monte Carlo method employed by the authors.}
\end{thebibliography}

\begin{widetext}

\appendix

\section{\label{sec:supp_mat}Supplemental Material for ``Topological Kondo
insulators in one dimension: Continuous Haldane-type ground-state
evolution from the strongly-interacting to the non-interacting limit''}

\author{Franco T. Lisandrini}

\affiliation{Facultad de Ciencias Exactas Ingenier\'ia y Agrimensura, Universidad
Nacional de Rosario, Argentina}

\affiliation{Instituto de F\'isica Rosario, CCT-Rosario (CONICET-UNR), Bv. 27
de Febrero 210 bis, 2000 Rosario, Argentina}

\author{Alejandro M. Lobos}

\email{lobos@ifir-conicet.gov.ar}

\selectlanguage{english}%

\affiliation{Instituto de F\'isica Rosario, CCT-Rosario (CONICET), Bv. 27 de
Febrero 210 bis, 2000 Rosario, Argentina}

\affiliation{Facultad de Ciencia Exactas y Naturales, Universidad Nacional de
Cuyo, 5500 Mendoza, Argentina }

\author{Ariel O. Dobry}

\affiliation{Facultad de Ciencias Exactas Ingenier\'ia y Agrimensura, Universidad
Nacional de Rosario, Argentina}

\affiliation{Instituto de F\'isica Rosario, CCT-Rosario (CONICET-UNR), Bv. 27
de Febrero 210 bis, 2000 Rosario, Argentina}

\author{Claudio J. Gazza}

\affiliation{Facultad de Ciencias Exactas Ingenier\'ia y Agrimensura, Universidad
Nacional de Rosario, Argentina}

\affiliation{Instituto de F\'isica Rosario, CCT-Rosario (CONICET-UNR), Bv. 27
de Febrero 210 bis, 2000 Rosario, Argentina}

\date{\today}

\section*{Calculation of the string-correlator for a non-interacting flat-band
$sp$ ladder}

In this Supplemental Material we give details on the calculation of
the string correlator $\mathcal{O}_{\text{string}}^{z}\left(l-m\right)$
in the non-interacting case $U=0$ and for flat-band parameters $t_{s}=t_{p}=t_{sp}=t$.
This analytical result is striking since, despite the absence of well-defined
magnetic moments in the system due to charge fluctuations, it shows
that time-reversal $\mathbb{Z}_{2}$topological band-insulators can
have a non-vanishing value of $\mathcal{O}_{\text{string}}^{z}\left(l-m\right)$,
a quantity typically associated to the strongly interacting Haldane
phase. 

As mentioned in the main text, in the flat-band case the Hamiltonian
reads%
\begin{align*}
H_{\text{flat}} & =t\sum_{j=1}^{L-1}\sum_{\sigma}\left[\left(s_{j\sigma}^{\dagger}-p_{j\sigma}^{\dagger}\right)\left(s_{j+1\sigma}+p_{j+1\sigma}\right)+\left(s_{j+1\sigma}^{\dagger}+p_{j+1\sigma}^{\dagger}\right)\left(s_{j\sigma}-p_{j\sigma}\right)\right]\\
 & =2t\sum_{j=1,\sigma}^{L-1}\left(\gamma_{-,j\sigma}^{\dagger}\gamma_{+,j+1\sigma}+\text{H.c.}\right),
\end{align*}
where 
\begin{align}
\gamma_{\pm j,\sigma} & =\frac{s_{j,\sigma}\pm p_{j,\sigma}}{\sqrt{2}}.\label{eq:gammaj}
\end{align}
This Hamiltonian can be diagonalized using the basis:

\begin{align}
B_{j,j+1,\sigma}^{\dagger} & =\frac{\gamma_{-,j\sigma}^{\dagger}+\gamma_{+,j+1\sigma}^{\dagger}}{\sqrt{2}},\label{eq:B}\\
A_{j,j+1,\sigma}^{\dagger} & =\frac{\gamma_{-,j\sigma}^{\dagger}-\gamma_{+,j+1\sigma}^{\dagger}}{\sqrt{2}},\label{eq:A}
\end{align}
which defines new ``sites'' along the diagonal rungs, and in terms
of which the Hamiltonian writes%
\begin{align*}
H_{\text{flat}}= & -2t\sum_{j=1,\sigma}^{L-1}\left(A_{j,j+1,\sigma}^{\dagger}A_{j,j+1,\sigma}-B_{j,j+1,\sigma}^{\dagger}B_{j,j+1,\sigma}\right).
\end{align*}

The groundstate of a system with $N=2L$ particles (i.e., half-filling),
where the boundary modes are occupied with fermions $\gamma_{+,1s}^{\dagger},\gamma_{-,Ls^{\prime}}^{\dagger}$
can be expressed as 
\begin{align}
\left|\psi_{0}\left(s,s^{\prime}\right)\right\rangle  & =\left\{ \prod_{j=1}^{L-1}\prod_{\sigma}A_{j,j+1,\sigma}^{\dagger}\right\} \gamma_{+,1s}^{\dagger}\gamma_{-,Ls^{\prime}}^{\dagger}\left|0\right\rangle .\label{eq:groundstate}
\end{align}
The contribution inside the curly brackets corresponds to the gapped
1D bulk, while the last two terms are the end-states. 

We now focus on the string correlator

\begin{align}
\mathcal{O}_{\text{string}}^{z}\left(l-m\right) & \equiv-\left\langle T_{l}^{z}e^{i\pi\sum_{l<j<m}T_{j}^{z}}T_{m}^{z}\right\rangle \label{eq:string_order}
\end{align}
where
\begin{align}
T_{j}^{z} & =\frac{n_{j,\uparrow}^{s}-n_{j,\downarrow}^{s}+n_{j,\uparrow}^{p}-n_{j,\downarrow}^{p}}{2}=\frac{n_{+,j\uparrow}-n_{+j,\downarrow}+n_{-,j\uparrow}-n_{-,j\downarrow}}{2},\label{eq:Tjz}
\end{align}
where we have defined $n_{\pm,j\sigma}=\gamma_{\pm,j\sigma}^{\dagger}\gamma_{\pm,j\sigma}$.
Since all the operators $T_{j}^{z}$ inside the expression $T_{l}^{z}e^{i\pi\sum_{l<j<m}T_{j}^{z}}T_{m}^{z}$
commute, we compute the average of $T_{l}^{z}T_{m}^{z}e^{i\pi\sum_{l<j<m}T_{j}^{z}}$
instead. Therefore, we first operate with the string on the ground
state $\left|\psi_{0}\left(s,s^{\prime}\right)\right\rangle $. To
that end it is useful to recall the commutation relation

\begin{align}
e^{i\alpha c^{\dagger}c}c^{\dagger} & =c^{\dagger}e^{i\alpha\left(c^{\dagger}c+1\right)},\label{eq:commutation}
\end{align}
which arises from the fermionic commutation relation $\left[c^{\dagger}c,c\right]=c^{\dagger}$.
Using this relationship, we compute the effect of the string on the
operator $A_{k,k+1,\sigma}$: %
\begin{align}
e^{i\pi\sum_{j=l+1}^{j=m-1}T_{j}^{z}}A_{k,k+1,\sigma} & =\left(\frac{\gamma_{-,k\sigma}^{\dagger}e^{i\sigma\frac{\pi}{2}\left[\Theta\left(k-l-1\right)-\Theta\left(k-m+1\right)\right]}-\gamma_{+,k+1\sigma}^{\dagger}e^{i\sigma\frac{\pi}{2}\left[\Theta\left(k-l\right)-\Theta\left(k-m+2\right)\right]}}{\sqrt{2}}\right)e^{i\pi\sum_{j=l+1}^{j=m-1}T_{j}^{z}},\label{eq:commutation_exp_A}
\end{align}
where we have defined the unit step function as:
\begin{align*}
\Theta\left(j\right) & =\begin{cases}
1 & \text{for }j\geq0,\\
0 & \text{for }j<0.
\end{cases}
\end{align*}
We see that the effect of the string is to introduce a phase factor
$e^{i\sigma\frac{\pi}{2}}$ on every creation operator $\gamma_{\pm,k\sigma}^{\dagger}$
which verifies $l+1\leq k\leq m-1$. When both operators $\gamma_{+,k\sigma}^{\dagger}$
and $\gamma_{-,k+1\sigma}^{\dagger}$ inside the definition (\ref{eq:A})
are affected by the string, the phase factors for spin $\sigma=\uparrow$
and $\sigma=\downarrow$ compensate and cancel out as a consequence
of the SU(2) symmetry of the ground state, and therefore the original
operator $A_{k,k+1,\sigma}$ is restored. However, note that for $A_{l,l+1,\sigma}$
and $A_{m-1,m,\sigma}$, this is not possible and therefore after
the action of the string, and a phase factor remains and we cannot
restore $A_{l,l+1,\sigma}$ nor $A_{m-1,m,\sigma}$. In the thermodynamic
limit $L\rightarrow\infty$, and assuming $\left|l-m\right|\geqslant2$
with $l$ and $m$ deep in the 1D bulk, we can neglect the effect
of the boundaries and therefore we can write
\begin{align}
e^{i\pi\sum_{j=l+1}^{j=m-1}T_{j}^{z}}\left|\psi_{0}\left(s,s^{\prime}\right)\right\rangle  & =\left\{ \prod_{\sigma}\left(\frac{\gamma_{-,l\sigma}^{\dagger}-e^{i\sigma\frac{\pi}{2}}\gamma_{+,l+1\sigma}^{\dagger}}{\sqrt{2}}\right)\left(\frac{e^{i\sigma\frac{\pi}{2}}\gamma_{-,m-1\sigma}^{\dagger}-\gamma_{+,m\sigma}^{\dagger}}{\sqrt{2}}\right)\right\} \nonumber \\
 & \times\left\{ \prod_{j\neq\left\{ l,m-1\right\} }^{L-1}\prod_{\sigma}A_{j,j+1,\sigma}^{\dagger}\right\} \gamma_{+,1s}^{\dagger}\gamma_{-,Ls^{\prime}}^{\dagger}\left|0\right\rangle ,\label{eq:string_operator_on_groundstate}
\end{align}
where in the first brackets we have singled out the operators which
cannot be restored into the original form $A_{k,k+1,\sigma}$, while
in the last brackets we have regrouped the rest of the unaffected
operators.

Now let us focus on the effect of $T_{l}^{z}T_{m}^{z}$. Since $T_{l}^{z}$
and $T_{m}^{z}$ are local operators acting on sites $l$ and $m$
respectively, the four operators $A_{l-1,l,\sigma}$,$A_{l,l+1,\sigma}$
and $A_{m-1,m,\sigma}$,$A_{m,m+1,\sigma}$, two of which have already
been modified by the string, will be affected. Remember that we are
assuming $\left|l-m\right|\geqslant2$, otherwise some of these operators
will coincide. It is convenient to single out these four terms and
write (\ref{eq:string_operator_on_groundstate}) as

\begin{align}
e^{i\pi\sum_{j=l+1}^{j=m-1}T_{j}^{z}}\left|\psi_{0}\left(s,s^{\prime}\right)\right\rangle  & =\left\{ \prod_{\sigma}\left(\frac{\gamma_{-,l-1\sigma}^{\dagger}-\gamma_{+,l\sigma}^{\dagger}}{\sqrt{2}}\right)\left(\frac{\gamma_{-,l\sigma}^{\dagger}-e^{i\sigma\frac{\pi}{2}}\gamma_{+,l+1\sigma}^{\dagger}}{\sqrt{2}}\right)\right.\\
 & \left.\times\left(\frac{e^{i\sigma\frac{\pi}{2}}\gamma_{-,m-1,\sigma}^{\dagger}-\gamma_{+,m\sigma}^{\dagger}}{\sqrt{2}}\right)\left(\frac{\gamma_{-,m\sigma}^{\dagger}-\gamma_{+,m+1\sigma}^{\dagger}}{\sqrt{2}}\right)\right\} \times\mathbf{A}_{s,s^{\prime}}^{\dagger}\left|0\right\rangle \label{eq:string_op_gs_compact}\\
 & =\left\{ \prod_{\sigma}\left(\frac{\gamma_{-,l-1,\sigma}^{\dagger}\gamma_{-,l\sigma}^{\dagger}-\gamma_{+,l,\sigma}^{\dagger}\gamma_{-,l\sigma}^{\dagger}-e^{i\sigma\frac{\pi}{2}}\gamma_{-,l-1,\sigma}^{\dagger}\gamma_{+,l+1,\sigma}^{\dagger}+e^{i\sigma\frac{\pi}{2}}\gamma_{+,l,\sigma}^{\dagger}\gamma_{+,l+1,\sigma}^{\dagger}}{2}\right)\right.\\
 & \left.\times\left(\frac{e^{i\sigma\frac{\pi}{2}}\gamma_{-,m-1,\sigma}^{\dagger}\gamma_{-,m\sigma}^{\dagger}-e^{i\sigma\frac{\pi}{2}}\gamma_{-,m-1,\sigma}^{\dagger}\gamma_{+,m+1\sigma}^{\dagger}-\gamma_{+,m\sigma}^{\dagger}\gamma_{-,m\sigma}^{\dagger}+\gamma_{+,m\sigma}^{\dagger}\gamma_{+,m+1\sigma}^{\dagger}}{2}\right)\right\} \mathbf{A}_{s,s^{\prime}}^{\dagger}\left|0\right\rangle 
\end{align}
where

\begin{align}
\mathbf{A}_{s,s^{\prime}}^{\dagger} & =\left\{ \prod_{j\neq\left\{ l-1,l,m-1,m\right\} }^{L-1}\prod_{\sigma}A_{j,j+1,\sigma}^{\dagger}\right\} \gamma_{+,1s}^{\dagger}\gamma_{-,Ls^{\prime}}^{\dagger},\label{eq:compact_non_affected}
\end{align}
is the operator which is not affected by neither the string nor by
the spin operators $T_{l}^{z}T_{m}^{z}$. Then, operating with $T_{l}^{z}T_{m}^{z}$
on the state $e^{i\pi\sum_{j=l+1}^{j=m-1}T_{j}^{z}}\left|\psi_{0}\left(s,s^{\prime}\right)\right\rangle $
and using Eq. (\ref{eq:Tjz}) yields%

\begin{align*}
\left|\phi_{l,m,s,s^{\prime}}\right\rangle  & \equiv T_{l}^{z}T_{m}^{z}e^{i\pi\sum_{j=l+1}^{j=m-1}T_{j}^{z}}\left|\psi_{0}\left(s,s^{\prime}\right)\right\rangle ,\\
 & =\frac{1}{4}\left\{ \left[\left(\frac{\gamma_{-,l-1,\uparrow}^{\dagger}\gamma_{-,l\uparrow}^{\dagger}-2\gamma_{+,l,\uparrow}^{\dagger}\gamma_{-,l\uparrow}^{\dagger}+e^{i\frac{\pi}{2}}\gamma_{+,l,\uparrow}^{\dagger}\gamma_{+,l+1,\uparrow}^{\dagger}}{2}\right)A_{l-1,l,\downarrow}^{\dagger}A_{l,l+1,\downarrow}^{\dagger}\right.\right.\\
 & \left.-A_{l-1,l,\uparrow}^{\dagger}A_{l,l+1,\uparrow}^{\dagger}\left(\frac{\gamma_{-,l-1,\downarrow}^{\dagger}\gamma_{-,l\downarrow}^{\dagger}-2\gamma_{+,l,\downarrow}^{\dagger}\gamma_{-,l\downarrow}^{\dagger}+e^{-i\frac{\pi}{2}}\gamma_{+,l,\downarrow}^{\dagger}\gamma_{+,l+1,\downarrow}^{\dagger}}{2}\right)\right]\\
 & \times\left[\left(\frac{e^{i\frac{\pi}{2}}\gamma_{-,m-1,\uparrow}^{\dagger}\gamma_{-,m\uparrow}^{\dagger}-2\gamma_{+,m\uparrow}^{\dagger}\gamma_{-,m\uparrow}^{\dagger}+\gamma_{+,m\uparrow}^{\dagger}\gamma_{+,m+1\uparrow}^{\dagger}}{2}\right)A_{m-1,m,\downarrow}^{\dagger}A_{m,m+1,\downarrow}^{\dagger}\right.\\
 & \left.\left.-A_{m-1,m,\uparrow}^{\dagger}A_{m,m+1,\uparrow}^{\dagger}\left(\frac{e^{-i\frac{\pi}{2}}\gamma_{-,m-1,\downarrow}^{\dagger}\gamma_{-,m\downarrow}^{\dagger}-2\gamma_{+,m\downarrow}^{\dagger}\gamma_{-,m\downarrow}^{\dagger}+\gamma_{+,m\downarrow}^{\dagger}\gamma_{+,m+1\downarrow}^{\dagger}}{2}\right)\right]\right\} \mathbf{A}_{s,s^{\prime}}^{\dagger}\left|0\right\rangle 
\end{align*}

Finally, we compute the string-order correlator as 

\begin{align*}
\mathcal{O}_{\text{string}}^{z}\left(l-m\right) & =-\left\langle \psi_{0}\left(s,s^{\prime}\right)\right|T_{l}^{z}e^{i\pi\sum_{l<j<m}T_{j}^{z}}T_{m}^{z}\left|\psi_{0}\left(s,s^{\prime}\right)\right\rangle ,\\
 & =-\left\langle \psi_{0}\left(s,s^{\prime}\right)|\phi_{l,m,s,s^{\prime}}\right\rangle .
\end{align*}
Using the fact that the operators $A_{j,j+1,\sigma},\ A_{j,j+1,\sigma}^{\dagger}$
anticommute with all other operators $A_{k,k+1,\sigma},\ A_{k,k+1,\sigma}^{\dagger}$
with $j\neq k$, it is easy to see that $A_{j,j+1,\sigma}$ and $A_{j,j+1,\sigma}^{\dagger}$
cancels out, since $\left(A_{j,j+1,\sigma}A_{j,j+1,\sigma}^{\dagger}\right)\left|0\right\rangle =\left|0\right\rangle $
. Then, projecting $\left|\phi_{l,m,s,s^{\prime}}\right\rangle $
onto $\left|\psi_{0}\left(s,s^{\prime}\right)\right\rangle $ yields%

\begin{align}
\mathcal{O}_{\text{string}}^{z}\left(l-m\right) & =-\left\langle \psi_{0}\left(s,s^{\prime}\right)|\phi_{l,m,s,s^{\prime}}\right\rangle ,\nonumber \\
 & =-\frac{1}{4}\left\langle 0\right|\left\{ \left[\left(A_{l,l+1,\uparrow}A_{l-1,l,\uparrow}\right)\left(\frac{\gamma_{-,l-1,\uparrow}^{\dagger}\gamma_{-,l\uparrow}^{\dagger}-2\gamma_{+,l,\uparrow}^{\dagger}\gamma_{-,l\uparrow}^{\dagger}+e^{i\frac{\pi}{2}}\gamma_{+,l,\uparrow}^{\dagger}\gamma_{+,l+1,\uparrow}^{\dagger}}{2}\right)\right.\right.\nonumber \\
 & \left.-\left(A_{l,l+1,\downarrow}A_{l-1,l,\downarrow}\right)\left(\frac{\gamma_{-,l-1,\downarrow}^{\dagger}\gamma_{-,l\downarrow}^{\dagger}-2\gamma_{+,l,\downarrow}^{\dagger}\gamma_{-,l\downarrow}^{\dagger}+e^{-i\frac{\pi}{2}}\gamma_{+,l,\downarrow}^{\dagger}\gamma_{+,l+1,\downarrow}^{\dagger}}{2}\right)\right]\nonumber \\
 & \times\left[\left(A_{m,m+1,\uparrow}A_{m-1,m,\uparrow}\right)\left(\frac{e^{i\frac{\pi}{2}}\gamma_{-,m-1,\uparrow}^{\dagger}\gamma_{-,m\uparrow}^{\dagger}-2\gamma_{+,m\uparrow}^{\dagger}\gamma_{-,m\uparrow}^{\dagger}+\gamma_{+,m\uparrow}^{\dagger}\gamma_{+,m+1\uparrow}^{\dagger}}{2}\right)\right.\nonumber \\
 & \left.\left.-\left(A_{m,m+1,\downarrow}A_{m-1,m,\downarrow}\right)\left(\frac{e^{-i\frac{\pi}{2}}\gamma_{-,m-1,\downarrow}^{\dagger}\gamma_{-,m\downarrow}^{\dagger}-2\gamma_{+,m\downarrow}^{\dagger}\gamma_{-,m\downarrow}^{\dagger}+\gamma_{+,m\downarrow}^{\dagger}\gamma_{+,m+1\downarrow}^{\dagger}}{2}\right)\right]\right\} \left|0\right\rangle ,\nonumber \\
 & =-\left\{ \frac{\left(3+e^{i\frac{\pi}{2}}\right)-\left(3+e^{-i\frac{\pi}{2}}\right)}{8}\right\} ^{2},\nonumber \\
 & =\frac{1}{16}\qquad(\text{for }\left|l-m\right|\geqslant2),\label{eq:str_order_final}
\end{align}
as mentioned in the main text. It is interesting to note that all
the effect of the string is encoded by the phase factors $e^{\pm i\frac{\pi}{2}}$.
Should the string was not there, i.e., if we compute the spin-spin
correlator 
\begin{align}
\mathcal{C}^{z}\left(l-m\right) & \equiv\left\langle \psi_{0}\left(s,s^{\prime}\right)\right|T_{l}^{z}T_{m}^{z}\left|\psi_{0}\left(s,s^{\prime}\right)\right\rangle \label{eq:correlation}
\end{align}
the only change would be to replace the factors $e^{\pm i\frac{\pi}{2}}\rightarrow1$
in the above Eq. (\ref{eq:str_order_final}) {[}see also Eq. (\ref{eq:commutation_exp_A}){]}.
In that case, we would obtain
\begin{align}
\mathcal{C}^{z}\left(l-m\right)=\left\{ \frac{\left(3+1\right)-\left(3+1\right)}{8}\right\} ^{2} & =0, & (\text{for }\left|l-m\right|\geqslant2)\label{eq:correlation_final}
\end{align}
as expected, since the system has no net magnetic moments.\end{widetext}
\end{document}